\documentstyle[epsfig]{mn}
\newif\ifAMStwofonts
\AMStwofontstrue

\ifoldfss
  \ifCUPmtlplainloaded \else
    \NewTextAlphabet{textbfit} {cmbxti10} {}
    \NewTextAlphabet{textbfss} {cmssbx10} {}
    \NewMathAlphabet{mathbfit} {cmbxti10} {} 
    \NewMathAlphabet{mathbfss} {cmssbx10} {} 
  \fi
  \ifAMStwofonts
    \ifCUPmtlplainloaded \else
      \NewSymbolFont{upmath} {eurm10}
      \NewSymbolFont{AMSa} {msam10}
      \NewMathSymbol{\upi}     {0}{upmath}{19}
      \NewMathSymbol{\umu}     {0}{upmath}{16}
      \NewMathSymbol{\upartial}{0}{upmath}{40}
      \NewMathSymbol{\leqslant}{3}{AMSa}{36}
      \NewMathSymbol{\geqslant}{3}{AMSa}{3E}

    \fi
  \fi
\fi 

\ifnfssone
  \newmathalphabet{\mathit}
  \addtoversion{normal}{\mathit}{cmr}{m}{it}
  \addtoversion{bold}{\mathit}{cmr}{bx}{it}
  \newmathalphabet{\mathbfit} 
  \addtoversion{normal}{\mathbfit}{cmr}{bx}{it}
  \addtoversion{bold}{\mathbfit}{cmr}{bx}{it}
  \newmathalphabet{\mathbfss} 
  \addtoversion{normal}{\mathbfss}{cmss}{bx}{n}
  \addtoversion{bold}{\mathbfss}{cmss}{bx}{n}
  \ifAMStwofonts
    \ifCUPmtlplainloaded \else
      \UseAMStwoboldmath
      \makeatletter
      \new@mathgroup\upmath@group
      \define@mathgroup\mv@normal\upmath@group{eur}{m}{n}
      \define@mathgroup\mv@bold\upmath@group{eur}{b}{n}
      \edef\UPM{\hexnumber\upmath@group}
      \new@mathgroup\amsa@group
      \define@mathgroup\mv@normal\amsa@group{msa}{m}{n}
      \define@mathgroup\mv@bold\amsa@group{msa}{m}{n}
      \edef\AMSa{\hexnumber\amsa@group}
      \makeatother
      \mathchardef\upi="0\UPM19
      \mathchardef\umu="0\UPM16
      \mathchardef\upartial="0\UPM40
      \mathchardef\leqslant="3\AMSa36
      \mathchardef\geqslant="3\AMSa3E
    \fi
  \fi
\fi 

\ifnfsstwo
  \DeclareMathAlphabet{\mathbfit}{OT1}{cmr}{bx}{it}
  \SetMathAlphabet\mathbfit{bold}{OT1}{cmr}{bx}{it}
  \DeclareMathAlphabet{\mathbfss}{OT1}{cmss}{bx}{n}
  \SetMathAlphabet\mathbfss{bold}{OT1}{cmss}{bx}{n}
  \ifAMStwofonts
    \ifCUPmtlplainloaded \else
      \DeclareSymbolFont{UPM}{U}{eur}{m}{n}
      \SetSymbolFont{UPM}{bold}{U}{eur}{b}{n}
      \DeclareSymbolFont{AMSa}{U}{msa}{m}{n}
      \DeclareMathSymbol{\upi}{0}{UPM}{"19}
      \DeclareMathSymbol{\umu}{0}{UPM}{"16}
      \DeclareMathSymbol{\upartial}{0}{UPM}{"40}
      \DeclareMathSymbol{\leqslant}{3}{AMSa}{"36}
      \DeclareMathSymbol{\geqslant}{3}{AMSa}{"3E}
    \fi
  \fi
\fi 

\ifCUPmtlplainloaded \else
  \ifAMStwofonts \else 
    \def\upi{\pi}
    \def\umu{\mu}
    \def\upartial{\partial}
  \fi
\fi
\title{No Higher Criticism of the Bianchi Corrected WMAP Data}
\author[] { 
L. Cay\'on$^{1}$, A.J. Banday$^{2}$, T. Jaffe$^{2}$, H.K.K. Eriksen$^{3,4,5}$, F.K. Hansen$^{3,4}$,\\
\newauthor K.M. Gorski$^{5}$ and J. Jin$^{6}$ \\
1. Department of Physics. Purdue University. 525 Northwestern Avenue, West Lafayette, IN 47907-2036.\\
2.Max-Planck-Institut f\"ur Astrophysik,
Karl-Schwarzschild-Str.\ 1, Postfach 1317, D-85741 Garching bei
M\"unchen, Germany\\
3.Institute of Theoretical Astrophysics, University of
Oslo, P.O.\ Box 1029 Blindern, N-0315 Oslo, Norway\\ 
4. Centre of
Mathematics for Applications, University of Oslo, P.O.\ Box 1053
Blindern, N-0316 Oslo\\
5. JPL, M/S 169/327, 4800 Oak Grove Drive,
Pasadena CA 91109; California Institute of Technology, Pasadena, CA
91125\\
6. Department of Statistics. Purdue University. 150 N. University Street, West Lafayette, IN 47907-2067.\\}

\date{\today}

\pagerange{\pageref{firstpage}--\pageref{lastpage}}
\pubyear{2001}

\begin{document}

\maketitle

\label{firstpage}

\begin{abstract}

\noindent  
Motivated by the success of the Bianchi $VII_h$ model in addressing 
many of the anomalies observed in the WMAP data (Jaffe et al.),
we present calculations in real and in wavelet space of the 
Higher Criticism statistic of the Bianchi corrected 
{\it Wilkinson Microwave Anisotropy Probe} (WMAP) first year data. 
At the wavelet scale of 5 degrees 
the Higher Criticism of the WMAP map drops from a value above the 
$99\%$ c.l. to a value below the $68\%$ 
c.l. when corrected by the Bianchi template. An important property of the 
Higher Criticism statistic is its ability to locate the pixels that
account for the deviation from Gaussianity. The analysis of the 
uncorrected WMAP data pointed to a cold spot in the southern 
hemisphere centered at $l\sim 209^{\circ}$, $b\sim -57^{\circ}$. 
The Higher Criticism of the Bianchi corrected map indicates that this
spot remains prominent, albeit at a level completely consistent with
Gaussian statistics. Consequently, it is debatable how much emphasis
should be placed on this residual feature, but we consider the effect
of modestly increasing the scaling of the template. A factor of only
1.2 renders the spot indistinguishable from the background level, with
no noticeable impact on the results published in Jaffe et al. for
the low-l anomalies, large-scale power asymmetry or wavelet kurtosis.
A trivial interpretation would be that the Bianchi template may require 
a small enhancement of power on scales corresponding to the wavelet scale
of 5 degrees.

\end{abstract}


\section{Introduction}

The first year of observations of the {\it Wilkinson Microwave Anisotropy 
Probe} (WMAP) satellite has provided us with a data set of unprecedented accuracy (Bennett et al. 2003). It is exciting to see that we are still far from 
understanding all the information that is encoded in the data. Among the many 
papers that have been published based on analyses of the WMAP data, a large
section of them points to possible deviations from the standard model. Different so called anomalies have been detected and possible explanations have been  
offered by several authors: (1) Low value of the quadrupole, its alignment with the octopole and other multipole alignments (De Oliveira-Costa et al. 2004, Schwarz et al. 2004, Vale 2005, Land \& Magueijo 2005a), (2) Asymmetries (Park 2004, Eriksen et al. 2004a, Eriksen et al. 2004b, 
Larson \& Wandelt 2004, Hansen et al. 2004a, Hansen et al. 2004b, Donoghue \& Donoghue 2005, Tojeiro et al. 2005, Bielewicz et al. 2005, Tomita 2005, Freeman et al. 2005), (3) Deviations from Gaussianity, in particular, detections in wavelet space pointing to a cold spot as the source of non-Gaussianity (Vielva et al. 2004, Cruz et al. 2005, Cay\'on et al. 2005). 

Several works have explored the possibility of accounting for some or all
of the anomalies observed in the WMAP data by introducing some shear and
vorticity in the universe (Jaffe et al. 2005a,b, Land \& Magueijo 2005b, McEwen et al. 2005b). Models introducing these anisotropic
characteristics fall under the class of Bianchi $VII_h$ models (Barrow et al. 1985). In this paper we confirm the validity of the Bianchi model 
obtained as a best fit to the WMAP data by Jaffe et al. 2005b, as a possible 
explanation of all the observed anomalies. 
Following the same procedure as in Cay\'on et al. 2005 we show that the Higher
Criticism $HC$ statistic (Donoho $\&$ Jin 2004) of the Bianchi corrected WMAP 
map is compatible with Gaussianity (at the $99\%$ c.l.) at all the 
wavelet scales.   

This paper is organized as follows. We present the formalism in Section 1. 
Section 2 is dedicated to introduce the analyzed data and simulations. 
Results are presented in Section 3 and they are discussed in Section 4 (conclusions also included in this section).

\section{Formalism}

As indicated above we here follow the formalism presented in Cay\'on et al. 2005. The statistical study is based on the $HC$ statistic proposed in 
Donoho \& Jin 2004, Jin 2004. For a set of $n$ individual observations $X_i$ 
from a certain distribution, $HC$ is defined as follows. The $X_i$ observed 
values are first converted into $p$-values: 
$p_i = P\{ |N(0,1)| > |X_i| \}$. After sorting the $p$-values in 
ascending order $p_{(1)} < p_{(2)} < \ldots < p_{(n)}$, we define the 
$HC$ at each pixel $i$ by: 
$$
HC_{n,i}  =    \sqrt{n} \biggl|  \frac{i/n  - p_{(i)}}{\sqrt{p_{(i)} (1-p_{(i)})}}  \biggr|,  
$$
Unusually large values of $HC_{n,i}$ imply deviations from Gaussianity. The fact that the statistic is calculated at every pixel will allow for the location
of the source of any detected deviation. 
In order to quantify the statistical level of any detection the maximum 
of the $HC_{n,i}$ for a given map (generally denoted $HC$) 
will be compared to the distribution of $HC$ values obtained from 
Gaussian simulations (see discussion in the next section). 
$HC$ has been shown to capture the unusual behavior of the few most
extreme observations as well as any unusually large number of 
moderately high value observations. One should note that this is 
a completely different test to one based on the kurtosis statistic. 
The kurtosis enhances any deviation from Gaussianity hidden in the 
fourth order moment and it depends on the bulk of the data.  

The $HC$ of the Bianchi corrected WMAP data is estimated in real and
in wavelet space.  Several groups have presented analyses of the 
WMAP data in wavelet space (Vielva et al. 2004, Cruz et al. 2005, Mukherjee \& Wang 2004, McEwen et al. 2005a,Liu \& Zhang 2005, Cay\'on et al. 2005).  
In particular, the Spherical Mexican Hat (SMH) wavelet has been shown to 
be sensitive to the presence of non-Gaussianity in the WMAP data. Analyses
based on kurtosis and on $HC$ show detections above the $99\%$ c.l. 
at a scale of five degrees. A cold spot in the 
southern hemisphere seems to be accounting for these detections. Our aim 
in this paper is to study what is the effect on the $HC$ results 
caused by the subtraction of a Bianchi based template from the WMAP data. 
$HC$ is therefore calculated not only for the Bianchi corrected 
WMAP map but also for the convolution of this map with 
the SMH wavelet at fifteen 
scales ranging from $13.74$ to $1050.0$ arcmins.

\setcounter{figure}{0}
\begin{figure*}
 \epsfxsize=84mm
 \epsffile{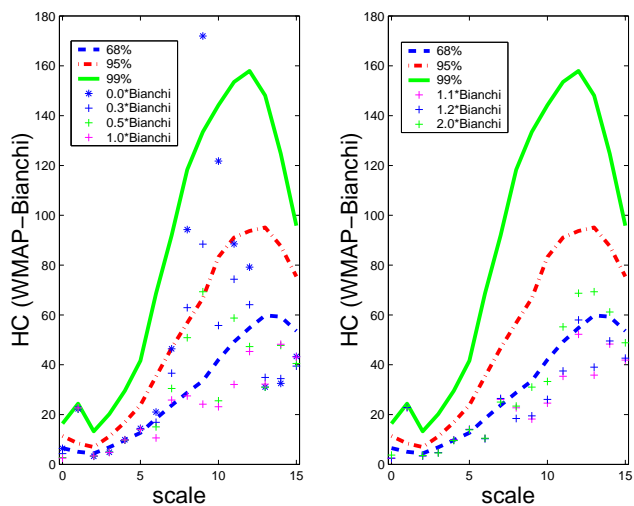}
 \caption{Values of the $HC$ statistic for the WMAP (stars) and the 
Bianchi corrected WMAP (crosses) data. The bands outlined
by dashed, dotted-dashed and solid lines correspond to the $68\%$, $95\%$ and $99\%$ confidence regions respectively. The Bianchi template is multiplied
by different factors before subtraction. Factors $0.3, 0.5$ and $1.0$ are
considered in the results presented in the left panel (blue, green and 
magenta crosses respectively). Results for factors $1.1, 1.2$ and $2.0$ are
presented in the right panel (magenta, blue and green crosses respectively).}
 \label{f1}
\end{figure*}

\section{Bianchi corrected WMAP Data and Simulations}

The first year of WMAP observations is available at the Legacy Archive for Microwave Background Data Analysis 
(LAMBDA) website\footnote[2]{http://lambda.gsfc.nasa.gov/}. For the purpose
of cosmological studies the Foreground Cleaned Intensity Maps at bands Q,V and W are used. A combined map (the so called WMAP data/map in this work) is built
following Komatsu et al. 2003. The weighted combination is given by,
$$T(i)={{\sum_{r=3}^{10} T_r(i)w_r(i)}\over {\sum_{r=3}^{10} w_r(i)}}.$$
The temperature at pixel $i$, $T(i)$ results from the ratio of the weighted sum 
of temperatures at pixel $i$ at each radiometer divided by the sum of the 
weights of each radiometer at pixel $i$. The radiometers Q1, Q2, V1, V2, W1, W2, W3 and W4 are sequentially numbered from 3 to 10. The weights at each pixel,
for each radiometer $w_r(i)$, are the ratio of the number of observations
$N_r(i)$ divided by the square of the receiver noise dispersion $\sigma_{o,r}$.
The resulting map is downgraded from resolution $nside=512$ to resolution
$nside=256$ (the total number of pixels being $12\times nside^2$).

The Bianchi template that is used in this work is the best 
fit found by Jaffe et al. 2005b. 
They considered Bianchi type $VII_h$ models
characterized by the values of $\Omega_o$ and 
$x=(h/(1-\Omega_o))^{1/2}$, where $h$ is the scale on which the basis
vectors change orientation. The best fit was obtained for $\Omega_o=0.5$ and
$x=0.62$. We present here an analysis of the Bianchi corrected
WMAP data set obtained after subtraction of 
this template (and multiples of it) from the original WMAP data.   
The final analysis is performed on a masked version of this map that 
includes the so called $Kp0$ mask (released by the WMAP team and available 
at the LAMBDA website). An extension of this mask is applied to the 
wavelet convolved maps as discussed in Vielva et al. 2004.

The confidence levels of the $HC$ statistic corresponding to the Gaussian 
assumption are drawn from 5000 simulations were the power spectrum is the one
that best fits the WMAP, CBI and ACBAR CMB data, plus the 2dF and Lyman-alpha
data. The simulations take also into account the beam transfer functions, the 
number of observations and the noise dispersion for each receiver. All these
are provided by the WMAP team through the LAMBDA website.

\setcounter{figure}{1}
\begin{figure*}
  \epsfig{file=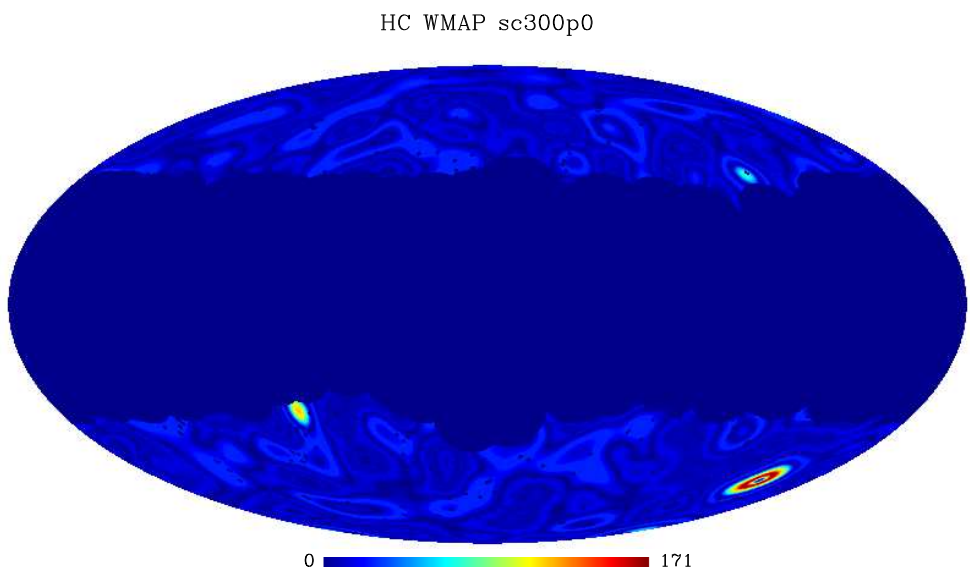,angle=90,scale=0.5}
  \caption{$HC$ values of the WMAP data at the wavelet scale of five degrees.}
 \label{f1}
\end{figure*}

\section{Results}

The values of the $HC$ statistic for the WMAP and the Bianchi corrected WMAP
data are presented in Figure 1. Solid, dotted-dashed and dashed lines 
show the $99\%, 95\%$ and $68\%$ confidence levels respectively. These are 
obtained from
the 5000 Gaussian simulations described above. Since the $HC$ statistic of
a map is a maximum value ($HC=max_i(HC_{n,i})$), 
the confidence
regions are one-sided. As one can see, there is no detection of
deviations from Gaussianity in the Bianchi corrected 
WMAP data (factor of $1.0$ in the figure).  This corroborates the results obtained by Jaffe et al. 2005b. Subtraction of the Bianchi template corrects 
most of the anomalies that have been observed in the first year
WMAP data.

The maps of $HC$ values for the WMAP and the Bianchi corrected 
WMAP data at a wavelet scale of 5 degrees are presented in Figures 2 and 3.
As one can immediately see the detected non-Gaussianity in the WMAP data
is dominated by the values of a ring of pixels at the spot centered at 
$l\sim 209^{\circ}$, $b\sim -57^{\circ}$ (note that the ring structure is 
likely to be caused by the convolution with the wavelet). Subtraction of the 
Bianchi template drops the values of the $HC$ statistic at that spot 
below the $68\%$ c.l. (see Figure 1). 
However, the spot still appears as one of the most
prominent regions. Even if the relevance of this feature 
is debatable we decided to see whether different normalizations 
of the subtracted template can influence its amplitude. 
A factor of $1.2$ reduces the amplitude of the spot making it 
comparable with the surrounding values as can be seen in Figure 4.
We have also studied whether this normalization factor affects the 
other anomalies observed in the WMAP first year data. The effect on the alignment between
the quadrupole and the octopole and the planarity as defined in Oliveira-Costa et al. 2004 is presented in Table 1. These large scale anomalies are not
very much affected by a change in the normalization factor of the 
Bianchi template. 
The ratio of power between two hemispheres 
for a certain range of $l$ was calculated as in Eriksen et al. 2004a. 
The probability of having a maximum power asymmetry ratio
larger than that of a given map (either the WMAP or the Bianchi corrected
WMAP maps) is shown in Table 2.  
This probability is slightly affected by the normalization 
factor. There is no much of a difference between the results for a factor
of 1 and a factor of $1.2$. However, above this, an increase 
in the normalization factor starts worsening the results. 
This agrees with the trend observed 
for the $HC$ statistic as can be seen in Figure 1.

\begin{table*}
  \begin{minipage}{170mm}
  \begin{center}
  \caption[]{Alignment and planarity of low order multipoles. The normalization $F$ of the Bianchi template is included in the first column. Second and third 
columns correspond to the direction of $l=2$ (Galactic longitude and latitude in degrees) . The direction of $l=3$ is shown in the fourth and fifth columns ($l$ and $b$ in degrees). The angle $\alpha$ (degrees) corresponding to the alignment between the quadrupole and the octopole is shown in column 6. The probability of finding a weaker alignment is presented in 
column 7. Values in the last three columns correspond to the probability of
finding a more planar multipole for $l=3,5,6$ respectively.}
  \label{tab1}
  \begin{tabular}{c|c|c|c|c|c|c|c|c|c}\hline
$F$  & $l_2$& $b_2$& $l_3$& $b_3$ & $\alpha$ & $P$& $P_3$&$P_5$&$P_6$\cr
  \hline\hline 

	0.0&61.9&247.8&63.4&232.8&7.0&0.992&0.89&0.001&0.98 \cr
	0.5&8.3&57.1&73.7 &253.3 &82.7 &0.127&0.73&0.003&0.99\cr
	1.0&3.3&44.5&56.8 &323.5 &82.3 &0.134&0.55&0.05&0.98\cr
	1.2&0.3&40.6&45.6 &327.2 &78.3 &0.203&0.62&0.10&0.97\cr
	2.0&9.0&212.2&26.9 &317.3&80.9 &0.159&0.86&0.37&0.92\cr

  \hline\hline
  \end{tabular}
  \end{center}
\end{minipage}
\end{table*}

\begin{table*}
  \begin{minipage}{170mm}
  \begin{center}
  \caption[]{Probability of the observed asymmetry in the power spectrum between the two hemispheres. The normalization factor $F$ is given in column 1. The
probability for $l=2,20$ is presented in column 2. Values in column 3 correspond
to the probability for $l=2,40$.}
  \label{tab1}
  \begin{tabular}{c|c|c}\hline
$F$  & $P(l=2,20)(\%)$& $P(l=2,40)(\%)$\cr
  \hline\hline 

	0.0&1.0&0.7\cr 
	0.5&9.8&6.0\cr 
	1.0&17.0&22.0\cr 
	1.2&15.9&28.9 \cr 
	2.0&4.3&13.6\cr

  \hline\hline
  \end{tabular}
  \end{center}
\end{minipage}
\end{table*}

\setcounter{figure}{2}
\begin{figure*}
  \epsfig{file=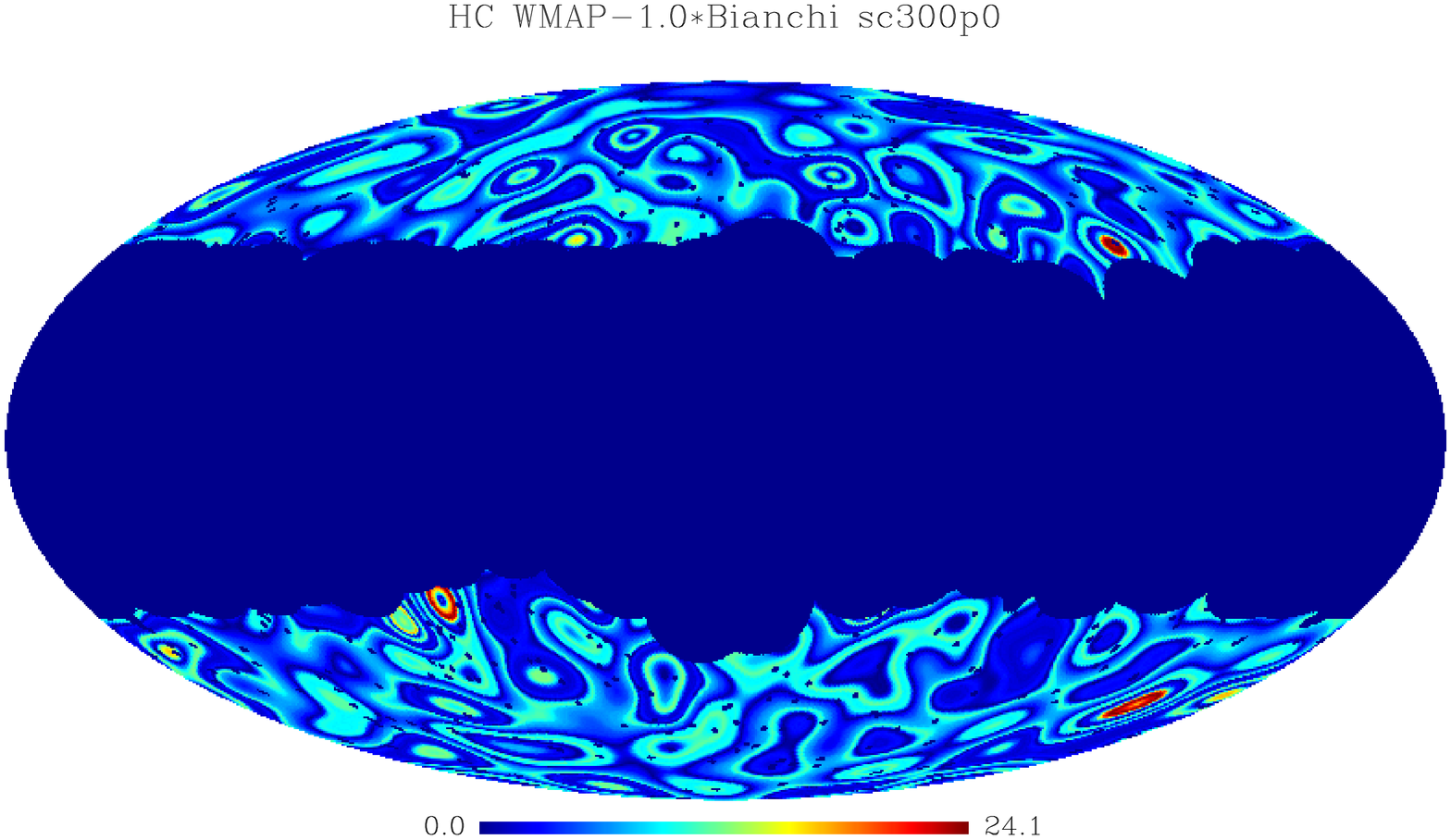,angle=90,scale=0.5}
  \caption{$HC$ values of the Bianchi corrected WMAP data at the wavelet scale of five degrees.}
 \label{f1}
\end{figure*}

\setcounter{figure}{3}
\begin{figure*}
  \epsfig{file=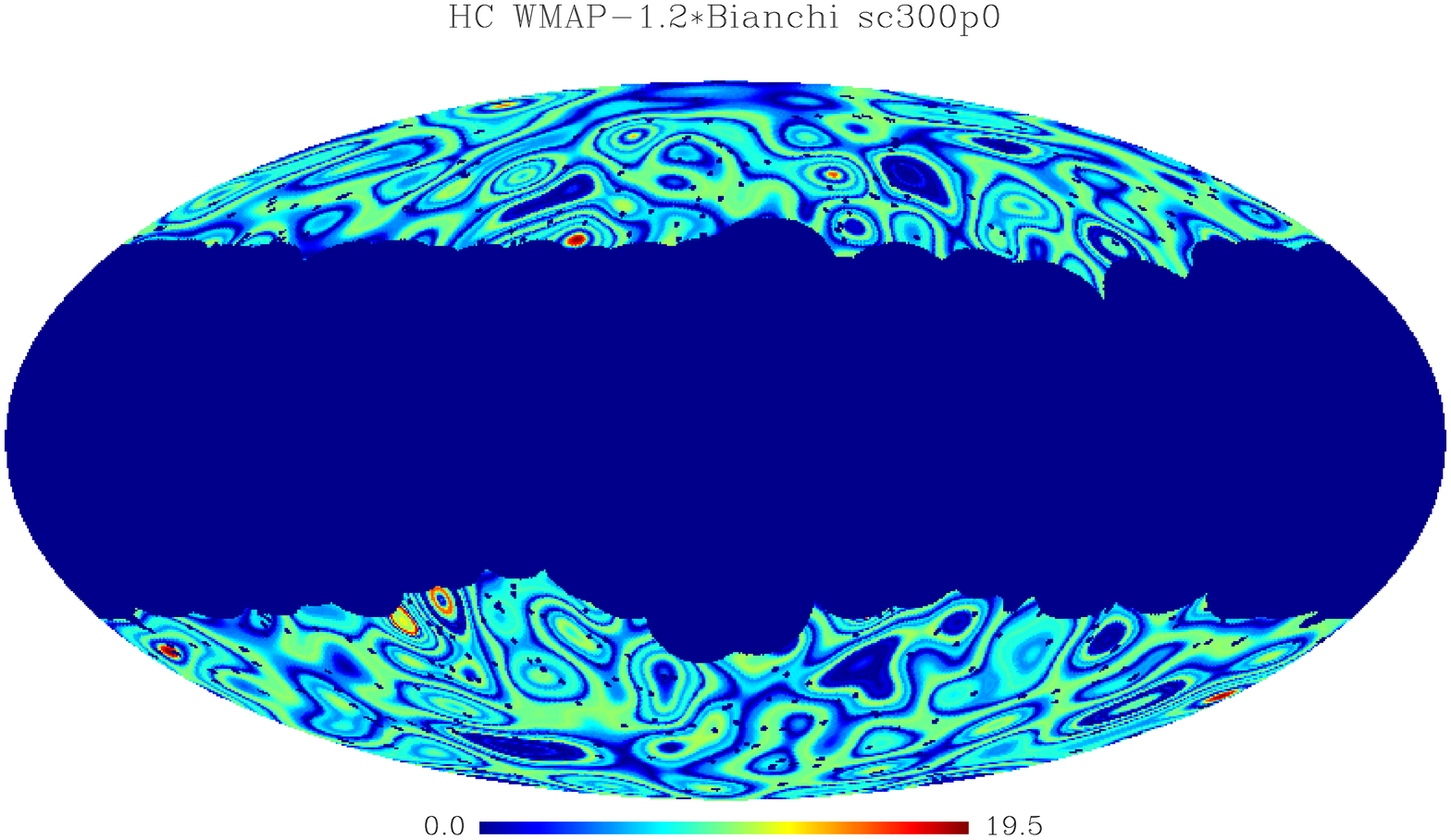,angle=90,scale=0.5}
  \caption{$HC$ values of the WMAP data minus $1.2$ times the 
Bianchi template, at the wavelet scale of five degrees.}
 \label{f1}
\end{figure*}

\section{Discussion and Conclusions}

Several explanations have been discussed in the literature to account
for the different anomalies observed in the first year of WMAP data (
Schwartz et al. 2004, Vale 2005, Freeman et al. 2005, Land \& Magueijo 2005b, Jaffe et al. 2005a,b, Tomita 2005). These anomalies include: (1) The low value
of the quadrupole, its alignment with the octopole and other multipole alignments, (2) 
The asymmetries observed in the distribution of
large scale power, (3) The non-Gaussianity detected in wavelet space which source seems to be centered at a cold spot in the southern hemisphere. However, 
much of this work has concentrated on explaining the 
anomalous low-l structure.
Jaffe et al, however, serendipitously discovered that at least part of 
the large angular scale structure of the WMAP data was described by a particular
Bianchi VIIh model, and that correcting the data for this contribution
provided a potential resolution for each of the above anomalies.
We here confirm their results by 
estimating the effect of subtracting the proposed Bianchi template from
the WMAP data, on the HC statistic. A detection of non-Gaussianity localized in a cold spot at $l\sim 209^{\circ}$, $b\sim -57^{\circ}$ was found by the 
estimation of the $HC$ statistic of the WMAP data (Cay\'on et al. 2005). 
Subtraction of the Bianchi template renders the WMAP data
statistically compatible with the expected levels based on Gaussian simulations.

The map of $HC$ values for the Bianchi corrected WMAP data (see Figure 3) still shows the 
cold spot in the southern hemisphere as one of the most prominent regions. 
Although no special consideration should now be given to the spot
on a statistical basis, it remains interesting given that the original
detection of non-Gaussianity was intimately associated with the region.
We have determined that a modest increase in the amplitude of the 
Bianchi template (by a factor ~ 1.2) renders the spot indistinguishable 
from the background level. 
Importantly, such a change in amplitude does not perturb the effect
of the template in resolving issues related to the low-l multipole 
features and power asymmetry.
There is no compelling statistical reason, therefore, to believe that
the Bianchi template amplitude derived in Jaffe et al should be considered
underestimated. The HC results presented here, however, may be interpreted
as implying that the Bianchi template may require 
a small enhancement of power on scales corresponding to the wavelet scale
of 5 degrees. One could create a
Bianchi template with more power on small scales by reducing the
matter density, but this would also affect the large-scale structure
and therefore the significance of the fit to the data.
The most pragmatic interpretation of 
Jaffe et al - that the best-fit Bianchi model provides a template temperature 
pattern which alternative models would need to reproduce in order to 
resolve the observed anomalous anisotropy structure - remains valid.

\section*{Acknowledgements}

We thank P. Vielva for providing the extended masks used at different
wavelet scales.
We acknowledge the use of the Legacy Archive for Microwave Background Data Analysis (LAMBDA). Support for LAMBDA is provided by the NASA Office of Space
Science. 
Some of the results in this paper have been derived using the HEALPix (G\'orski, Hivon, and Wandelt 1999) package.

\end{document}